\documentclass[12pt]{article}
\usepackage{epsf}
\title{ {\bf Lepton flavor conserving Z boson decays and scalar unparticle}}
\author{\vspace{1cm}\\
        {\bf E. O. Iltan,}
        \thanks{E-mail address:
        eiltan@newton.physics.metu.edu.tr}
 \\
        Physics Department, Middle East Technical University \\
        Ankara, Turkey\\}

\date{}

\begin{document}
\setlength{\baselineskip}{24pt}
\maketitle
\setlength{\baselineskip}{7mm}
\begin{abstract}
We predict the contribution of scalar unparticle to the branching
ratios of the lepton flavor conserving $Z\rightarrow l^+ l^-$
decays and we study the discrepancy between the experimental and
the QED corrected standard model branching ratios . We observe
that these decays are sensitive to the unparticle scaling
dimension $d_u$ for its small values, especially for heavy lepton
flavor output.
\end{abstract}
\thispagestyle{empty}
\newpage
\setcounter{page}{1}
%
Theoretically, Z boson decays to lepton pairs exist in the tree
level, in the standard model (SM) if the lepton flavor is
conserved. The improved experimental measurements stimulate the
studies of these interactions and with the Giga-Z option of the
Tesla project, there is a possibility to increase Z bosons at
resonance \cite{Hawkings}. The experimental predictions for the
branching ratios (BRs) of these decays are \cite{PartData}
\begin{eqnarray}
BR(Z\rightarrow e^+ e^-) &=& 3.363 \pm 0.004\,\%
\nonumber \, , \\
BR(Z\rightarrow \mu^+ \mu^-) &=& 3.366 \pm 0.007\,\%
\nonumber \, , \\
BR(Z\rightarrow \tau^+ \tau^-) &=& 3.370 \pm 0.0023 \,\% \, ,
\label{Expr1}
\end{eqnarray}
and the tree level SM predictions, including QED corrections read
\begin{eqnarray}
BR(Z\rightarrow e^+ e^-) &=& 3.3346\,\%
\nonumber \, , \\
BR(Z\rightarrow \mu^+ \mu^-) &=& 3.3346\,\%
\nonumber \, , \\
BR(Z\rightarrow \tau^+ \tau^-) &=& 3.3338 \,\% \, . \label{Expr2}
\end{eqnarray}
It is seen that the main contribution to BRs of Z boson lepton
pair decays is coming from the tree level SM contribution and the
discrepancy between the experimental and the SM results is of the
order of $1.0\,\%$. In the literature, there are various
experimental and theoretical studies
\cite{Kamon}-\cite{IltanZllNC}. The vector and axial coupling
constants in Z-decays have been measured at LEP \cite{LEP} and
various additional types of interactions have been performed. A
way to measure these contributions in the process $Z\rightarrow
\tau^+ \tau^-$ was described in \cite{Stiegler}. In
\cite{IltanZllFC} and \cite{IltanZllNC} the possible new physics
effects to the process $Z\rightarrow l^+ l^-$, in the two Higgs
doublet model and in the SM with the non-commutative effects have
been studied, respectively.

The present work is devoted to analysis whether the inclusion of
the scalar unparticle effects overcomes the discrepancy of the BRs
between the experimental and the QED corrected SM result (see
\cite{okun} and references therein) for the lepton flavor
conserving (LFC) Z decays. Furthermore, we study the new
parameters arising with the unparticle effects and the
dependencies of the BRs to these new parameters.

The unparticle idea is introduced by Georgi \cite{Georgi1,
Georgi2} and its effect in the processes, which are induced at
least in one loop level, is studied in various works
\cite{Lu}-\cite{IltanUn3}. This idea is based on the interaction
of the SM and the ultraviolet sector with non-trivial infrared
fixed point, at high energy level. The unparticles, being massless
and having non integral scaling dimension $d_u$, are new degrees
of freedom arising from the ultraviolet sector  around
$\Lambda_U\sim 1\,TeV$. The effective lagrangian which is
responsible for the interactions of unparticles with the SM fields
in the low energy level reads
\begin{equation}
{\cal{L}}_{eff}\sim
\frac{\eta}{\Lambda_U^{d_u+d_{SM}-n}}\,O_{SM}\, O_{U} \,,
\label{efflag}
\end{equation}
where $O_U$ is the unparticle operator, the parameter $\eta$ is
related to the energy scale of ultraviolet sector, the low energy
one and the matching coefficient \cite{Georgi1,Georgi2,Zwicky} and
$n$ is the space-time dimension.

\begin{figure}[htb]
\vskip 2.2truein \centering \epsfxsize=6.5in
\leavevmode\epsffile{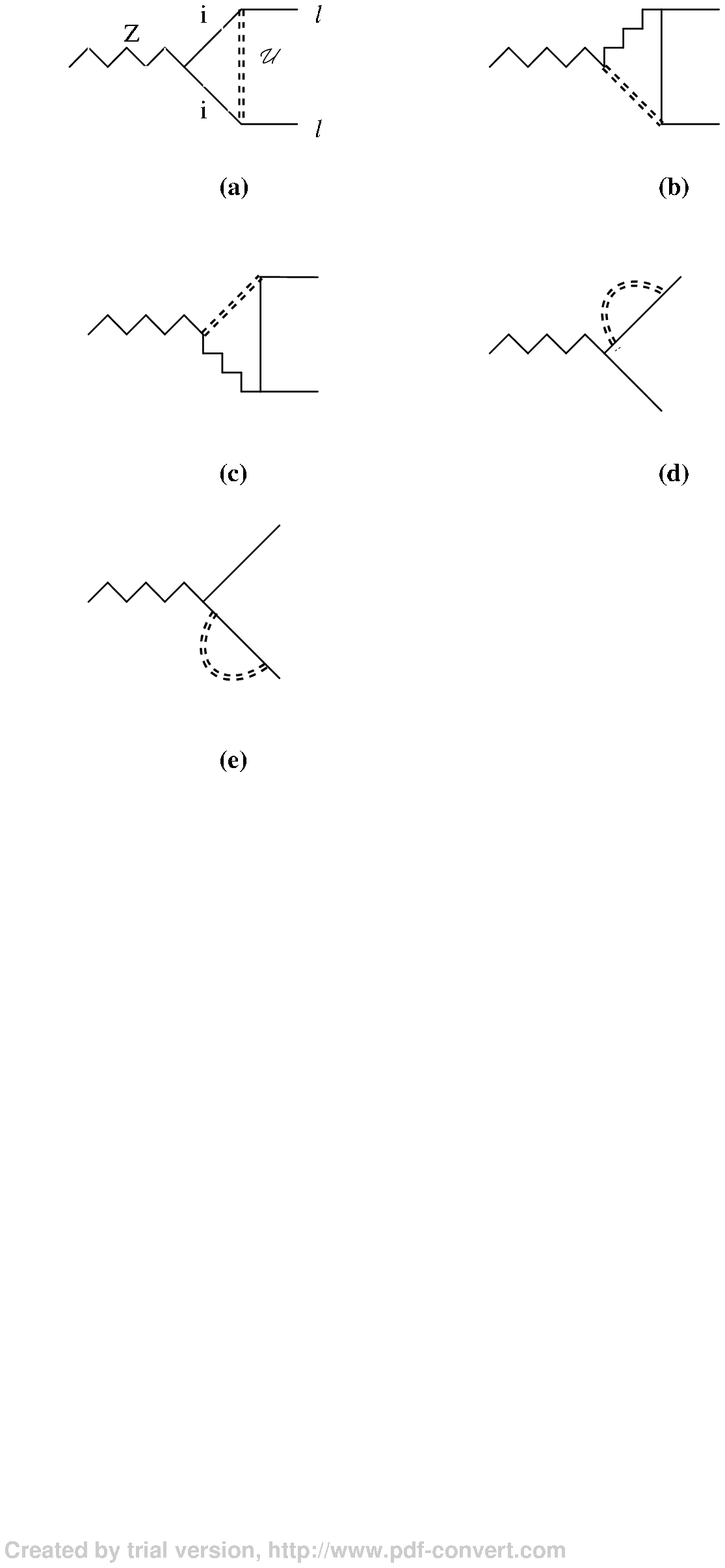} \vskip -5.5truein
\caption[]{One loop diagrams contribute to $Z\rightarrow l^+\,l^-$
decay with scalar unparticle mediator. Solid line represents the
lepton field: $i$ represents the internal lepton, $l^-$ ($l^+$)
outgoing lepton (anti lepton), wavy line the Z boson field, double
dashed line the unparticle field.} \label{figselfvert}
\end{figure}

Now, we present the effective lagrangian which drives the
$Z\rightarrow l^+ l^-$ decays with internal scalar unparticle
mediation. Here, we consider the operators with the lowest
possible dimension  since they have the most powerful effect in
the low energy effective theory (see for example \cite{SChen}).
The low energy effective interaction lagrangian which induces
$\textit{U}-l-l$ vertex  is
\begin{eqnarray}
{\cal{L}}_1= \frac{1}{\Lambda_U^{du-1}}\Big (\lambda_{ij}^{S}\,
\bar{l}_{i} \,l_{j}+\lambda_{ij}^{P}\,\bar{l}_{i}
\,i\gamma_5\,l_{j}\Big)\, O_{U} \, , \label{lagrangianscalar}
\end{eqnarray}
where $l$ is the lepton field and $\lambda_{ij}^{S}$
($\lambda_{ij}^{P}$) is the scalar (pseudoscalar)  coupling. In
addition to this lagrangian, the one which causes the tree level
$\textit{U}-Z-Z$ interaction  (see Fig \ref{figselfvert} (b) and
(c)), appearing in the scalar unparticle mediating loop, can exist
and it reads
\begin{eqnarray}
{\cal{L}}_2= \frac{\lambda_0}{\Lambda_U^{du}}\,
F_{\mu\nu}\,F^{\mu\nu}\, O_{U}+
\frac{\lambda_Z}{\Lambda_U^{du}}\, m_Z^2 Z^\mu\,Z_\mu \, O_{U}
\, , \label{lagrangianZ}
\end{eqnarray}
where $F_{\mu\nu}$ is the field tensor for the $Z_{\mu}$ field and
$\lambda_0$ and $\lambda_Z$ are effective coupling
constants\footnote{The vertex factor:
$\frac{i}{\Lambda_U^{d_u}}\,m_Z^2\,\lambda_Z\,g_{\mu\,\nu}+
\frac{4\,i}{\Lambda_U^{d_u}}\,\lambda_0\,
(k_{1\nu}\,k_{2\mu}-k_1.k_2\,g_{\mu\,\nu})$
where $k_{1(2)}$ is the four momentum of Z boson with polarization
vector $\epsilon_{1\,\mu \,(2\,\nu)}$.}.

Since the scalar unparticle contribution $Z\rightarrow l^+\,l^- $
decay enters into calculations at least in the one loop level (see
Fig.\ref{figselfvert}), one needs the scalar unparticle propagator
and it is obtained by using the scale invariance \cite{Georgi2,
Cheung1}:
\begin{eqnarray}
\!\!\! \int\,d^4x\,
e^{ipx}\,<0|T\Big(O_U(x)\,O_U(0)\Big)0>=i\frac{A_{d_u}}{2\,\pi}\,
\int_0^{\infty}\,ds\,\frac{s^{d_u-2}}{p^2-s+i\epsilon}=i\,\frac{A_{d_u}}
{2\,sin\,(d_u\pi)}\,(-p^2-i\epsilon)^{d_u-2} \, ,
\label{propagator}
\end{eqnarray}
where the function $\frac{1}{(-p^2-i\epsilon)^{2-d_u}}$ reads
\begin{eqnarray}
\frac{1}{(-p^2-i\epsilon)^{2-d_u}}\rightarrow
\frac{e^{-i\,d_u\,\pi}}{(p^2)^{2-d_u}} \, , \label{strongphase}
\end{eqnarray}
for $p^2>0$ and a non-trivial phase appears as a result of
non-integral scaling dimension. Here where the factor $A_{d_u}$ is
\begin{eqnarray}
A_{d_u}=\frac{16\,\pi^{5/2}}{(2\,\pi)^{2\,d_u}}\,
\frac{\Gamma(d_u+\frac{1}{2})} {\Gamma(d_u-1)\,\Gamma(2\,d_u)} \,
. \label{Adu}
\end{eqnarray}

At this stage, we are ready to consider the general effective
vertex for the interaction of on-shell Z-boson with a fermionic
current:
\begin{eqnarray}
\Gamma_{\mu}=\gamma_{\mu}(f_V-f_A\ \gamma_5)+
\frac{i}{m_W}\,(f_M+f_E\, \gamma_5)\, \sigma_{\mu\,\nu}\, q^{\nu}
\, , \label{vertex}
\end{eqnarray}
where $q$ is the momentum transfer, $q^2=(p-p')^2$, $f_V$ ($f_A$)
is vector (axial-vector) coupling, $f_M$ ($f_E$) is proportional
to the weak magnetic (electric dipole) moments of the fermion.
Here $p$ ($-p^{\prime}$) is the four momentum vector of lepton
(anti-lepton). The form factors $f_V$, $f_A$, $f_{M}$ and $f_{E}$
in eq. (\ref{vertex}) are obtained as
\begin{eqnarray}
f_V&=&f^{SM}_V+\int^{1}_{0}\,dx\,f^U_{V\,self}+\int^{1}_{0}\,dx\,\int^{1-x}_{0}\,dy\,
f^U_{V\,vert} \, ,\nonumber \\
f_A&=&f^{SM}_A+\int^{1}_{0}\,dx\,f^U_{A\,self}+\int^{1}_{0}\,dx\,\int^{1-x}_{0}\,dy\,
f^U_{A\,vert} \, ,\nonumber \\
f_M&=&\int^{1}_{0}\,dx\,\int^{1-x}_{0}\,dy\, f^U_{M\,vert} \, ,\nonumber \\
f_E&=&\int^{1}_{0}\,dx\,\int^{1-x}_{0}\,dy\,f^U_{E\,vert} \, ,
\label{funpart}
\end{eqnarray}
where the QED corrected\footnote{The corrections are taken to the
lowest approximation in $\alpha_{EM}$} SM form factors $f^{SM}_V$
and $f^{SM}_A$ are \cite{okun}
\begin{eqnarray}
f^{SM}_V&=&\frac{-i\,e}{c_W\,s_W}\,(\bar{c}_1+\bar{c}_2)\, ,\nonumber \\
f^{SM}_A&=& \frac{-i\,e}{c_W\,s_W}\,(\bar{c}_2-\bar{c}_1)\, ,
\label{funpart2}
\end{eqnarray}
with
\begin{eqnarray}
\bar{c}_1&=&c_1+\frac{3}{16}\,\Bigg(
\frac{\alpha_{EM}}{\pi}\,(2\,s_W^2-1)+\frac{4\,m_l^2}{m_Z^2}\Bigg)
\, ,\nonumber \\
\bar{c}_2&=& c_2+\frac{3}{8}\,\Bigg(
\frac{\alpha_{EM}}{\pi}\,s_W^2-\frac{2\,m_l^2}{m_Z^2}\Bigg)\, .
\label{funpart3}
\end{eqnarray}
Here the parameters $c_1$ and $c_2$ read
\begin{eqnarray}
c_1&=&-\frac{1}{2}+s_W^2 \, ,\nonumber \\
c_2&=&s_W^2\, . \label{c12}
\end{eqnarray}
On the other hand the explicit expressions of the form factors
$f^U_{V\,self}$, $f^U_{A\,self}$, $f^U_{V\,vert}$,
$f^U_{A\,vert}$, $f^U_{M\,vert}$ and $f^U_{E\,vert}$, carrying
scalar unparticle effects, are
\begin{eqnarray}
f^U_{V\,self}&=& \frac{-i\,c_{self}\,
(c_1+c_2)\,L_{self}^{d_u-1}\,(1-x)^{2-d_u}}{32\,s_W\,c_W\,(d_u-1)\,\pi^2}\,
\sum_{i=1}^3\, \Big((\lambda_{il}^S)^2+(\lambda_{il}^P)^2\Big)
\nonumber \, ,
\end{eqnarray}
\begin{eqnarray}
f^U_{A\,self}&=& \frac{i\,c_{self}\,
(c_2-c_1)\,L_{self}^{d_u-1}\,(1-x)^{2-d_u}}{32\,s_W\,c_W\,(d_u-1)\,\pi^2}\,
\sum_{i=1}^3\, \Big((\lambda_{il}^S)^2+(\lambda_{il}^P)^2\Big)
\nonumber \, ,
\end{eqnarray}
\begin{eqnarray}
f^U_{V\,vert}&=& \frac{i\,
c_{ver}\,(c_1+c_2)\,(1-x-y)^{1-d_u}}{32\,\pi^2}\,
\sum_{i=1}^3\,\frac{1}{\,L_{vert}^{2-d_u}}\,
\Bigg\{2\,((\lambda_{il}^S)^2-(\lambda_{il}^P)^2)\,m_i\,
m_{l}\,(1-x-y)\nonumber \\  &+&
((\lambda_{il}^S)^2+(\lambda_{il}^P)^2)\,\Bigg (
m^2_i+m_Z^2\,x\,y+m_l^2\,(1-x-y)^2-\frac{L_{vert}}{1-d_u}\Bigg)
\Bigg\}
\nonumber \\ &+& \frac{\lambda_0\, m_Z^2}{16\,\pi^2}\,
\sum_{i=1}^3\,\Bigg\{
\frac{b_{ver}\,y^{1-d_u}}{L_{1\,vert}^{2-d_u}}\, \Big\{
m_i\,\Big((c_1-c_2)\,\lambda_{il}^P+i\,(c_1+c_2)\,\lambda_{il}^S\Big)
\,(x-y-1)\nonumber \\ &+&
m_l\,\Big((c_1-c_2)\,\lambda_{il}^P-i\,(c_1+c_2)\,\lambda_{il}^S\Big)
\,\Big((x+y)^2+y-x\Big) \Big\}
\nonumber \\ &+&
\frac{b^\prime_{ver}\,x^{1-d_u}}{L_{2\,vert}^{2-d_u}}\, \Big\{
m_i\,\Big((c_1-c_2)\,\lambda_{il}^P-i\,(c_1+c_2)\,\lambda_{il}^S\Big)
\,(x-y+1)\nonumber \\ &-&
m_l\,\Big((c_1-c_2)\,\lambda_{il}^P+i\,(c_1+c_2)\,\lambda_{il}^S\Big)
\,\Big((x+y)^2-y+x\Big) \Big\}\,\Bigg\}
 \nonumber \\
&+& \frac{\lambda_Z}{32\,\pi^2}\, \sum_{i=1}^3\,\Bigg\{
\frac{b_{ver}\,y^{1-d_u}}{L_{1\,vert}^{2-d_u}}\, \Big\{
\Big((c_1-c_2)\,\lambda_{il}^P-i\,(c_1+c_2)\,\lambda_{il}^S\Big)\,\Bigg(
m_Z^2\,m_l\,\Big(x\,y\,(x+y-1)\nonumber \\&+& x+y\Big) -
m_l^3\,(1-x-y)^2\,(x+y)+\frac{L_{1\,vert}}{2\,(d_u-1)}\,m_l\,
\Big(1+6(x+y-1)\Bigg)\nonumber \\&+&
\Big((c_1-c_2)\,\lambda_{il}^P+i\,(c_1+c_2)\,\lambda_{il}^S\Big)\,
m_i\,\Big(m_l^2\,(1-x-y)^2-m_Z^2-\frac{L_{1\,vert}}{2\,(d_u-1)}
\Big)\Big\}\nonumber \\ &+&
\frac{b^\prime_{ver}\,x^{1-d_u}}{L_{1\,vert}^{2-d_u}}\, \Big\{
\Big((c_1-c_2)\,\lambda_{il}^P+i\,(c_1+c_2)\,\lambda_{il}^S\Big)\,\Bigg(
m_l^3\,(1-x-y)^2\,(x+y)\nonumber
\\ &-& m_Z^2\,m_l\,\Big(x\,y\,(x+y-1)+x+y\Big) -
\frac{L_{2\,vert}}{2\,(d_u-1)}\,m_l\,
\Big(1+6(x+y-1)\Big)\nonumber \\&-&
\Big((c_1-c_2)\,\lambda_{il}^P-i\,(c_1+c_2)\,\lambda_{il}^S\Big)\,
m_i\,\Big(m_l^2\,(1-x-y)^2-m_Z^2-\frac{L_{2\,vert}}{2\,(d_u-1)}\Big)\Big\}
\Bigg\} \nonumber \, ,
\end{eqnarray}
%
\begin{eqnarray}
f^U_{A\,vert}&=& \frac{-i\,
c_{ver}\,(c_1-c_2)\,(1-x-y)^{1-d_u}}{32\,\pi^2}\,
\sum_{i=1}^3\,\frac{1}{\,L_{vert}^{2-d_u}}\,
\Bigg\{((\lambda_{il}^P)^2-(\lambda_{il}^S)^2)\,\Big (2\, m_i\,
m_{l}\,(1-x-y)\nonumber \\  &-&
((\lambda_{il}^S)^2+(\lambda_{il}^P)^2)\,\Bigg (
m^2_i-m_Z^2\,x\,y+m_l^2\,(1-x-y)^2-\frac{L_{vert}}{d_u-1}\Bigg)
\Bigg\}
\nonumber \\ &-& \frac{\lambda_0\, m_Z^2}{16\,\pi^2}\,
\sum_{i=1}^3\,\Bigg\{
\frac{b_{ver}\,y^{1-d_u}}{L_{1\,vert}^{2-d_u}}\, \Big\{
m_i\,\Big((c_1+c_2)\,\lambda_{il}^P+i\,(c_1-c_2)\,\lambda_{il}^S\Big)
\,(1-x+y)\nonumber \\ &-&
m_l\,\Big((c_1+c_2)\,\lambda_{il}^P-i\,(c_1-c_2)\,\lambda_{il}^S\Big)
\,\Big((x+y)\,(1-x+y)\Big) \Big\}
\nonumber \\ &+&
\frac{b^\prime_{ver}\,x^{1-d_u}}{L_{2\,vert}^{2-d_u}}\, \Big\{
m_i\,\Big((c_1+c_2)\,\lambda_{il}^P-i\,(c_1-c_2)\,\lambda_{il}^S\Big)
\,(y-x-1)\nonumber \\ &+&
m_l\,\Big((c_1+c_2)\,\lambda_{il}^P+i\,(c_1-c_2)\,\lambda_{il}^S\Big)
\,\Big((x+y)\,(1+x-y\Big) \Big\}\,\Bigg\}
 \nonumber \\
&+&\frac{\lambda_Z}{32\,\pi^2}\, \sum_{i=1}^3\,\Bigg\{
\frac{b_{ver}\,y^{1-d_u}}{L_{1\,vert}^{2-d_u}}\, \Big\{
\Big((c_1+c_2)\,\lambda_{il}^P-i\,(c_1-c_2)\,\lambda_{il}^S\Big)\,\Bigg(
m_Z^2\,m_l\,(x+y)\nonumber \\&+&
\frac{L_{1\,vert}}{2\,(d_u-1)}\,m_l \Bigg)-
\Big((c_1+c_2)\,\lambda_{il}^P+i\,(c_1-c_2)\,\lambda_{il}^S\Big)\,
m_i\,\Big(\frac{L_{1\,vert}}{2\,(d_u-1)}+m_Z^2
\Big)\Big\}\nonumber
\\ &-&
\frac{b^\prime_{ver}\,x^{1-d_u}}{L_{2\,vert}^{2-d_u}}\, \Big\{
\Big((c_1+c_2)\,\lambda_{il}^P+i\,(c_1-c_2)\,\lambda_{il}^S\Big)\,\Bigg(
m_Z^2\,m_l\,(x+y)\nonumber \\&+&
\frac{L_{2\,vert}}{2\,(d_u-1)}\,m_l \Bigg)-
\Big((c_1+c_2)\,\lambda_{il}^P-i\,(c_1-c_2)\,\lambda_{il}^S\Big)\,
m_i\,\Big(\frac{L_{1\,vert}}{2\,(d_u-1)}+m_Z^2 \Big)\Big\}\Bigg\}
\, ,
\end{eqnarray}
%
%
\begin{eqnarray}
f^U_{M\,vert}&=& -\frac{i\,(1-x-y)^{1-d_u}}{32\,\pi^2}\,
\sum_{i=1}^3\,\frac{c_{ver}\,m_Z\,c_W}{L_{vert}^{2-d_u}}\,
\Bigg\{m_i\, \Big( ((\lambda_{il}^S)^2-(\lambda_{il}^P)^2) \,
(c_1+c_2)\,(x+y)\nonumber \\&-& 2\,i\,
\lambda_{il}^S\,\lambda_{il}^P\,
(c_2-c_1)\,(x-y)\Big)+((\lambda_{il}^S)^2+(\lambda_{il}^P)^2)
\,(c_1+c_2)\,(1-x-y)\,(x+y)\Bigg\}
\nonumber \\ &-& \frac{i\,\lambda_0}{8\,\pi^2}\,
\sum_{i=1}^3\,\Bigg\{
\frac{b_{ver}\,m_Z\,c_W\,y^{1-d_u}}{L_{1\,vert}^{2-d_u}}\,\Big(
c_1\,(\lambda_{il}^S+i\,\lambda_{il}^P)+c_2\,
(\lambda_{il}^S-i\,\lambda_{il_2}^P)\Big)\,
\Big(m^2_Z\,x\,y+\frac{L_{1\,vert}}{d_u-1}\Big)
\nonumber \\ &+&
\frac{b^\prime_{ver}\,m_Z\,c_W\,x^{1-d_u}}{L_{2\,vert}^{2-d_u}}\,\Big(
c_1\,(\lambda_{il}^S-i\,\lambda_{il}^P)+c_2\,
(\lambda_{il}^S+i\,\lambda_{il_2}^P)\Big)\,
\Big(m^2_Z\,x\,y+\frac{L_{2\,vert}}{d_u-1}\Big)
 \Bigg\} \nonumber \\
&-&\frac{i\,\lambda_Z}{64\,\pi^2}\, \sum_{i=1}^3\,\Bigg\{
\frac{b_{ver}\,m_Z\,c_W\,y^{1-d_u}}{L_{1\,vert}^{2-d_u}}\, \Big\{
(c_1+c_2)\, \lambda_{il}^S\,\Big(-m_i\,m_l\,(1-x-y)^2 \nonumber
\\ &-&m_l^2\,(1-x-y)^2\,(x+y)+ m_Z^2\,x\, \Big(
2-y\,(1-x-y)\Big)+(3\,x+3\,y-2)\,\frac{L_{1\,vert}}{d_u-1} \Big)
\nonumber \\ &+& i\,(c_1-c_2)\,
\lambda_{il}^P\,\Big(m_i\,m_l\,(1-x-y)^2-
m_l^2\,(1-x-y)^2\,(x+y)\nonumber
\\ &+& m_Z^2\,x\, \Big(
2-y\,(1-x-y)\Big)+(3\,x+3\,y-2)\,\frac{L_{1\,vert}}{d_u-1} \Big)
\Big\}\nonumber \\
&+&
\frac{b^\prime_{ver}\,m_Z\,c_W\,x^{1-d_u}}{L_{2\,vert}^{2-d_u}}\,
\Big\{ (c_1+c_2)\,
\lambda_{il}^S\,\Big(-m_i\,m_l\,(1-x-y)^2-m_l^2\,(1-x-y)^2\,(x+y)\nonumber
\\ &+& m_Z^2\,y\, \Big(
2-x\,(1-x-y)\Big)+(3\,x+3\,y-2)\,\frac{L_{2\,vert}}{d_u-1} \Big) -
i\,(c_1-c_2)\, \lambda_{il}^P\,\Big(m_i\,m_l\,(1-x-y)^2\nonumber
\\ &-& m_l^2\,(1-x-y)^2\,(x+y)+m_Z^2\,y\, \Big(
2-x\,(1-x-y)\Big)+(3\,x+3\,y-2)\,\frac{L_{2\,vert}}{d_u-1} \Big)
\Big\} \Bigg\} \nonumber \, ,
\end{eqnarray}
\begin{eqnarray}
f^U_{E\,vert}&=& -\frac{i\,(1-x-y)^{1-d_u}}{32\,\pi^2}\,
\sum_{i=1}^3\,\frac{c_{ver}\,m_Z\,c_W}{L_{vert}^{2-d_u}}\,
\Bigg\{m_i\, \Big( ((\lambda_{il}^S)^2-(\lambda_{il}^P)^2) \,
(c_1-c_2)\,(x-y)\nonumber
\\&+& 2\,i\,\lambda_{il}^S\,\lambda_{il}^P\,(c_1+c_2)\,(x+y) \Big)+ m_l
((\lambda_{il}^S)^2+(\lambda_{il}^P)^2)\,
(c_1-c_2)\,(1-x-y)\,(x-y)
 \Bigg\}
\nonumber \\ &-& \frac{i\,\lambda_0}{8\,\pi^2}\,
\sum_{i=1}^3\,\Bigg\{
\frac{b_{ver}\,m_Z\,c_W\,y^{1-d_u}}{L_{1\,vert}^{2-d_u}}\,\Bigg(
\Big(c_1\,(\lambda_{il}^S+i\,\lambda_{il}^P)-c_2\,(\lambda_{il}^S-i\,
\lambda_{il}^P)\Big)\,\Big(m_Z^2\,x\,y\nonumber \\
&+& m_l^2\,(1-x-y)\,(x+y)-\frac{L_{1\,vert}}{1-d_u}\Big)
+\Big(c_1\,(\lambda_{il}^S-i\,\lambda_{il}^P)-c_2\,(\lambda_{il}^S+i\,
\lambda_{il}^P)\Big)\, m_i\,m_l\, (1-x-y)\Bigg)
\nonumber \\ &+&
\frac{b^\prime_{ver}\,m_Z\,c_W\,x^{1-d_u}}{L_{2\,vert}^{2-d_u}}\,\Bigg(
\Big(c_2\,(\lambda_{il}^S+i\,
\lambda_{il}^P)-c_1\,(\lambda_{il}^S-i\,\lambda_{il}^P)\Big)\,
\Big(m_Z^2\,x\,y+m_l^2\,(1-x-y)\,(x+y) \nonumber \\ &-&
\frac{L_{2\,vert}}{1-d_u}\Big) -
\Big(c_1\,(\lambda_{il}^S+i\,\lambda_{il}^P)-c_2\,(\lambda_{il}^S-i\,
\lambda_{il}^P)\Big)
\,m_i\,m_l\,(1-x-y)\Bigg) \Bigg\} \nonumber \\
&-& \frac{i\,\lambda_Z}{64\,\pi^2}\, \sum_{i=1}^3\,\Bigg\{
\frac{b_{ver}\,m_Z\,c_W\,y^{1-d_u}}{L_{1\,vert}^{2-d_u}}\, \Big\{
(i\, (c_1+c_2)\,
\lambda_{il}^P\,\Bigg(m_i\,m_l\,(y^2-(1-x)^2)\nonumber
\\ &+& m_l^2\,(1-x-y)\, ((x+y)^2-x+y)+ m_Z^2\,x\, \Big(
2-y\,(1-x-y)\Big)+(3\,x+3\,y-2)\,\frac{L_{1\,vert}}{d_u-1} \Bigg)
\nonumber\\ &-&(c_1-c_2)\, \lambda_{il}^S\,\Bigg(m_i\,m_l\,
(y^2-(1-x)^2)- m_l^2\,(1-x-y)\,((x+y)^2-x+y)\nonumber
\\ &-& m_Z^2\,x\, \Big( 2-y\,(1-x-y)\Big)-
(3\,x+3\,y-2\Big)\,\frac{L_{1\,vert}}{d_u-1} \Bigg)
\Big \} \nonumber \\
&+&
\frac{b^\prime_{ver}\,m_Z\,c_W\,x^{1-d_u}}{L_{2\,vert}^{2-d_u}}\,
\Big\{ i\,(c_1+c_2)\, \lambda_{il}^P\,\Bigg(
m_i\,m_l\,(x^2-(1-y)^2)\nonumber
\\ &+& m_l^2\,(1-x-y)\, ((x+y)^2-y+x)+ m_Z^2\,y\, \Big(
2-x\,(1-x-y)\Big)+(3\,x+3\,y-2)\,\frac{L_{2\,vert}}{d_u-1} \Bigg)
\nonumber \\ &+& (c_1-c_2)\, \lambda_{il}^S\,\Bigg(m_i\,m_l\,
(x^2-(1-y)^2)-m_l^2\,(1-x-y)\,((x+y)^2-y+x) \nonumber
\\ &-&  m_Z^2\,x\, \Big( 2-x\,(1-x-y)\Big)-(3\,x+3\,y-2)\,
\frac{L_{2\,vert}}{d_u-1} \Bigg) \Big\} \Bigg\} \, , \label{fAVME}
\end{eqnarray}
%
with
\begin{eqnarray}
L_{self}&=&x\,\Big(m_{l}^2\,(1-x)-m_i^2\Big)
\, , \nonumber \\
L_{vert}&=&m_{l}^2\,(x+y)\,(1-x-y)-m_i^2\,(x+y)+m_Z^2\,x\,y
\, , \nonumber \\
L_{1\,vert}&=&
\Big(m_{l}^2\,(x+y)-m_i^2\Big)\,(1-x-y)+m_Z^2\,x\,(y-1)
\, , \nonumber \\
L_{2\,vert}&=&
\Big(m_{l}^2\,(x+y)-m_i^2\Big)\,(1-x-y)+m_Z^2\,y\,(x-1) \label{Ll}
\, ,
\end{eqnarray}
and
\begin{eqnarray}
c_{self}&=&-\frac{e\,A_{d_u}}{2\,sin\,(d_u\pi)\,\Lambda_u^{2\,(d_u-1)}}\,
, \nonumber \\
c_{ver}&=&-\frac{e\,A_{d_u}}{2\,s_W\,c_W\,\,sin\,(d_u\pi)\,\Lambda_u^{2\,(d_u-1)}}
\, , \nonumber \\
b_{ver}&=&-\frac{e\,A_{d_u}}{2\,s_W\,c_W\,\,sin\,(d_u\pi)\,\Lambda_u^{2\,d_u-1}}
\, , \nonumber \\
b^\prime_{ver}&=&-b_{ver} \label{cselfver} \, .
\end{eqnarray}
In eq. (\ref{fAVME}), the flavor diagonal and flavor changing
scalar and pseudoscalar couplings $\lambda_{il}^{S,P}$ represent
the effective interaction between the internal lepton $i$,
($i=e,\mu,\tau$) and the outgoing $l^-\,(l^+)$ lepton (anti
lepton). Finally, using the form factors $f_V$, $f_A$, $f_M$ and
$f_E$, the BR for $Z\rightarrow l^-\,l^+$ decay is obtained as
\begin{eqnarray}
BR (Z\rightarrow l^+\,l^-)=\frac{1}{48\,\pi}\,
\frac{m_Z}{\Gamma_Z}\, \{|f_V|^2+|f_A|^2+\frac{1}{2\,c^2_W}
(|f_M|^2+|f_E|^2) \} \label{BR1} \, ,
\end{eqnarray}
where
$\Gamma_Z$ is the total decay width of Z boson.
%
\\ \\
{\Large \textbf{Discussion}}
\\ \\
This section is devoted to the scalar unparticle effect on the BRs
of LFC Z boson decays. LFC Z boson decays exist in the tree level
in the framework of the SM and there are discrepancies between the
SM BRs and the experimental ones. Here, we include the possible
scalar unparticle contribution, which appears at least in the one
loop, and search whether these contributions could explain the
discrepancies in the BRs. We also study the new free parameters
which appear with the inclusion of scalar unparticle contribution:
the scaling dimension $d_u$, the new couplings, the energy scale.
These parameters should be restricted by respecting the current
experimental measurements and some theoretical considerations.
First, we choose the scaling dimension $d_u$ in the
range\footnote{Here, $d_u>1$ is due to the non-integrable
singularities in the decay rate \cite{Georgi2} and $d_u<2$ is due
to the convergence of the integrals \cite{Liao1}.} $1< d_u <2$.
The scalar unparticles appear in the loops with the following new
couplings in the framework of the effective theory: the
$\textit{U}-l-l$ couplings $\lambda_{ij}$, the $\textit{U}-Z-Z$
couplings $\lambda_0$, $\lambda_Z$ (see eqs.
(\ref{lagrangianscalar}, \ref{lagrangianZ}) and Fig.
\ref{figselfvert}). For the $\textit{U}-l-l$ couplings we consider
that the diagonal ones $\lambda_{ii}$ are aware of flavor,
$\lambda_{\tau\tau}>\lambda_{\mu\mu}>\lambda_{ee}$ and the off
diagonal couplings $\lambda_{ij}$ are flavor blind,
$\lambda_{ij}=\kappa \lambda_{ee}$ with $\kappa < 1$. In our
numerical calculations, we choose $\kappa=0.5$. On the other hand,
the possible tree level $\textit{U}-Z-Z$ interaction (see eqs.
(\ref{lagrangianZ})) is induced by new couplings $\lambda_0$ and
$\lambda_Z$ (see eq. (\ref{lagrangianZ})) and, for these
couplings, we choose the range $0.1-1.0$. Finally, we take the
energy scale of the order of TeV.
Notice that throughout our calculations we use the input values
given in Table (\ref{input}).
\begin{table}[h]
        \begin{center}
        \begin{tabular}{|l|l|}
        \hline
        \multicolumn{1}{|c|}{Parameter} &
                \multicolumn{1}{|c|}{Value}     \\
        \hline \hline
        $m_e$           & $0.0005$   (GeV)  \\
        $m_{\mu}$                   & $0.106$ (GeV) \\
        $m_{\tau}$                  & $1.780$ (GeV) \\
        $\Gamma^{Tot}_Z$           & $2.49$ (GeV) \\
        $s_W^2$             & $0.23$  \\
        $\alpha_{EM}$             & $1/129$  \\
        $BR_{SM}(Z\rightarrow ee)$             & $0.03346$  \\
        $BR_{SM}(Z\rightarrow \mu\mu)$             & $0.03346$  \\
        $BR_{SM}(Z\rightarrow \tau\tau)$             & $0.03338$  \\
        \hline
        \end{tabular}
        \end{center}
\caption{The values of the input parameters used in the numerical
          calculations.}
\label{input}
\end{table}
\\

\begin{figure}[htb]
\vskip -3.0truein \centering \epsfxsize=6.8in
\leavevmode\epsffile{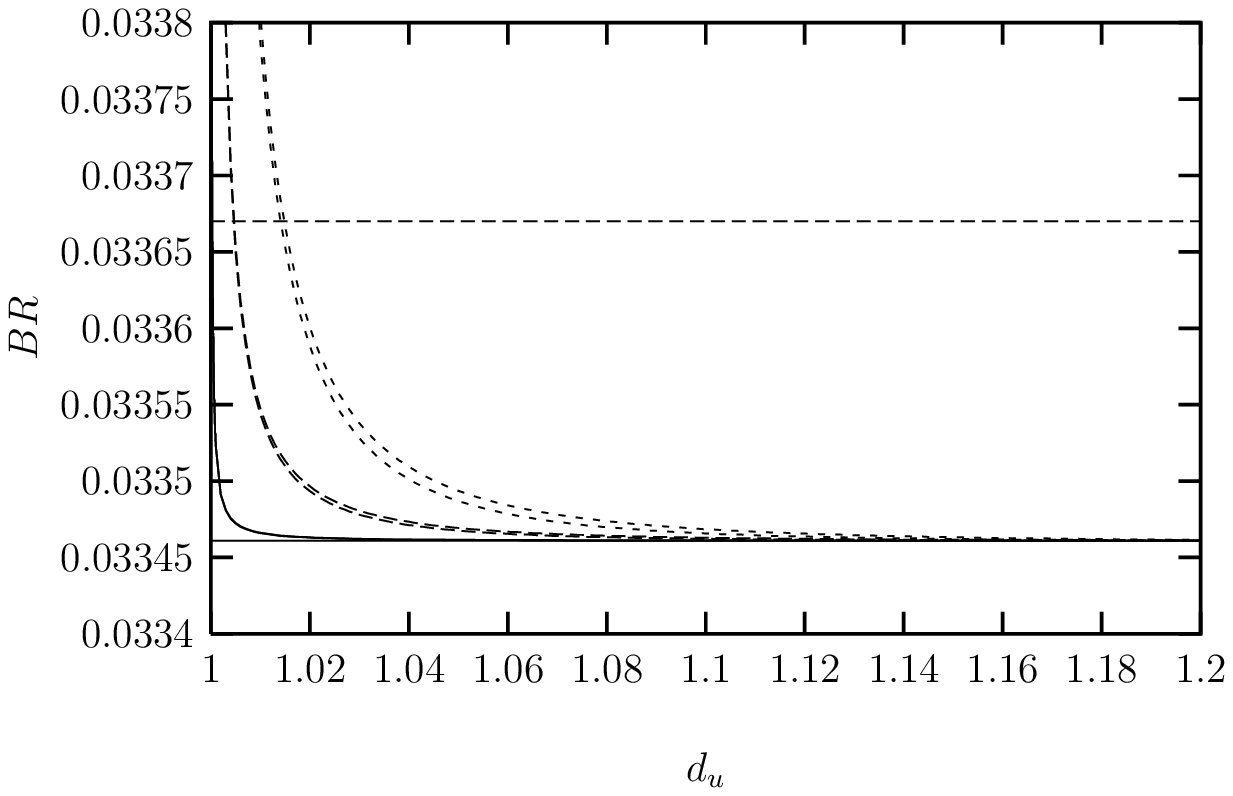} \vskip -3.0truein \caption[]{ The
scale parameter $d_u$ dependence of the BR $(Z\rightarrow e^+\,
e^-)$ for $\Lambda_u=10\, TeV$, $\lambda_{ee}=0.01$,
$\lambda_{\mu\mu}=0.1$, $\lambda_{\tau\tau}=1$,
$\lambda_{ij}=0.005$, $i\neq j$ and $\lambda_0=\lambda_Z=0.1$. The
solid (dashed) straight line represents the SM (experimental) BR
and the left-right solid (dashed, short dashed) curves represent
the BR including the scalar unparticle contribution, for the
energy scale $\Lambda_u=10\, TeV$-$\Lambda_u=1.0\, TeV$
$\lambda_{e}=0.01\,(0.05,\, 0.1)$.} \label{Zeedu}
\end{figure}

Fig. \ref{Zeedu} represents the BR $(Z\rightarrow e^+\, e^-)$ with
respect to the scale parameter $d_u$, for the couplings
$\lambda_{\mu\mu}=0.1$, $\lambda_{\tau\tau}=1$,
$\lambda_{ij}=0.5\,\lambda_{ee}$, $i\neq j$ and
$\lambda_0=\lambda_Z=0.1$. Here the solid (dashed) straight line
represents the QED corrected SM (the experimental\footnote{For the
experimental values of the BRs we use the numerical values which
are obtained by adding the experimental uncertainties to the mean
values.}) BR. On the other hand the left-right solid\footnote{The
solid lines almost coincide.} (dashed, short dashed) curves
represent the BR including the scalar unparticle contribution, for
the energy scale $\Lambda_u=10\, TeV$-$\Lambda_u=1.0\, TeV$,
$\lambda_{ee}=0.01\,(0.05,\, 0.1)$. The BR is sensitive to the
scale $d_u$ for its values near to one and the experimental result
is obtained in the case that the parameter $d_u$ has the values
$d_u\leq 1.02$, for the numerical values of the coupling
$\lambda_{ee}\sim 0.1$. The scalar unparticle contribution to the
BR is negligible for larger $d_u$ values.

\begin{figure}[htb]
\vskip -3.0truein \centering \epsfxsize=6.8in
\leavevmode\epsffile{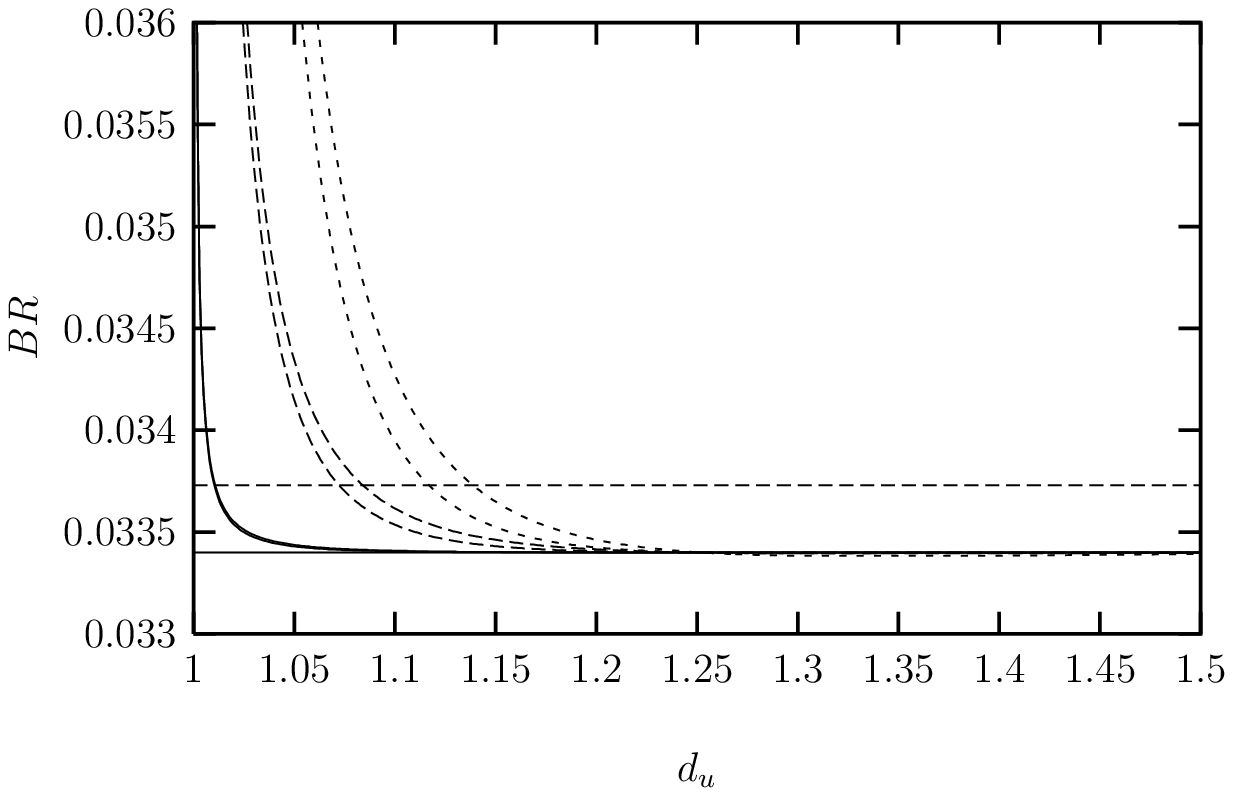} \vskip -3.0truein \caption[]{ The
scale parameter $d_u$ dependence of the BR $(Z\rightarrow \mu^+\,
\mu^-)$ for  $\lambda_{ee}=0.01$, $\lambda_{\tau\tau}=1$,
$\lambda_{ij}=0.005$, $i\neq j$ and $\lambda_0=\lambda_Z=0.1$. The
solid (dashed) straight line represents the SM (experimental) BR
and the left-right solid (dashed, short dashed) curves represent
the BR including the scalar unparticle contribution, for the
energy scale $\Lambda_u=10\, TeV$-$\Lambda_u=1.0\, TeV$
$\lambda_{\mu\mu}=0.1\,(0.5,\,1.0)$.} \label{Zmumudu}
\end{figure}

Fig. \ref{Zmumudu} shows the BR $(Z\rightarrow \mu^+\, \mu^-)$
with respect to the scale parameter $d_u$, for the couplings
$\lambda_{ee}=0.01$, $\lambda_{\tau\tau}=1$, $\lambda_{ij}=0.005$,
$i\neq j$ and $\lambda_0=\lambda_Z=0.1$. Here the solid (dashed)
straight line represents the QED corrected SM (the experimental)
BR and the left-right solid\footnote{The solid lines almost
coincide.} (dashed, short dashed) curves represent the BR
including the scalar unparticle contribution, for the energy scale
$\Lambda_u=10\, TeV$-$\Lambda_u=1.0\, TeV$
$\lambda_{\mu\mu}=0.1\,(0.5,\,1.0)$. Similar to the $Z\rightarrow
e^+\, e^-$ decay the BR is sensitive to the scale $d_u$ for its
values near to one and the experimental result is obtained for the
range of the parameter $d_u$, $d_u\leq 1.15$, for the numerical
values of the coupling $\lambda_{\mu\mu}\sim 1.0$. The BR is not
sensitive the scalar unparticle contribution for larger values of
$d_u$.

\begin{figure}[htb]
\vskip -3.0truein \centering \epsfxsize=6.8in
\leavevmode\epsffile{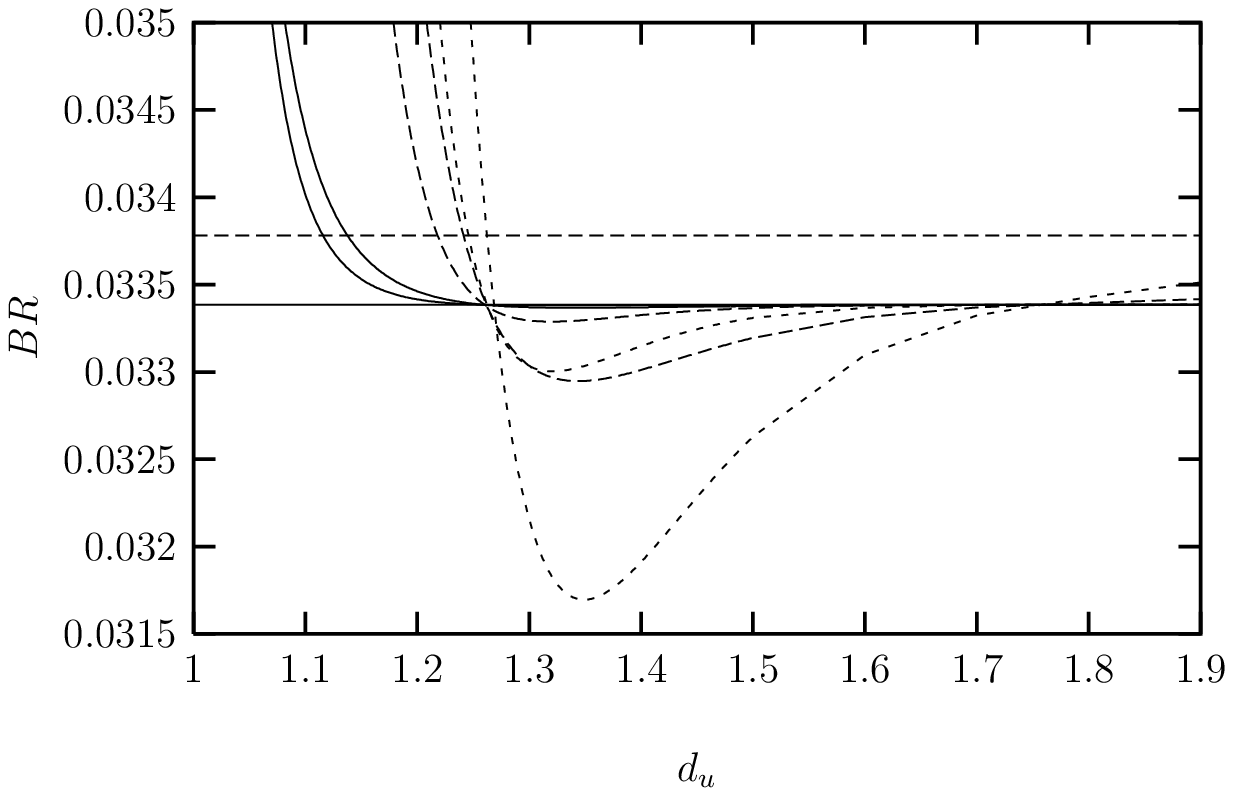} \vskip -3.0truein \caption[]{
The scale parameter $d_u$ dependence of the BR $(Z\rightarrow
\tau^+\, \tau^-)$ for $\lambda_{ee}=0.01$, $\lambda_{\mu\mu}=0.1$,
$\lambda_{ij}=0.005$, $i\neq j$ and $\lambda_0=\lambda_Z=0.1$. The
solid (dashed) straight line represents the SM (experimental) BR
and the left-right solid (dashed, short dashed) curves represent
the BR including the scalar unparticle contribution, for the
energy scale $\Lambda_u=10\, TeV$-$\Lambda_u=1.0\, TeV$
$\lambda_{\tau\tau}=1.0\,(5.0,\, 10)$.} \label{Ztautaudu}
\end{figure}

In Fig. \ref{Ztautaudu}, we present the BR $(Z\rightarrow \tau^+\,
\tau^-)$ with respect to the scale parameter $d_u$, for the
couplings $\lambda_{ee}=0.01$, $\lambda_{\mu\mu}=0.1$,
$\lambda_{ij}=0.005$, $i\neq j$ and $\lambda_0=\lambda_Z=0.1$.
Here the solid (dashed) straight line represents the QED corrected
SM (the experimental) BR and the left-right solid (dashed, short
dashed) curves represent the BR including the scalar unparticle
contribution, for the energy scale $\Lambda_u=10\,
TeV$-$\Lambda_u=1.0\, TeV$ $\lambda_{\tau\tau}=1.0\,(5.0,\, 10)$.
The addition of the scalar unparticle effect causes that the  BR
reaches to the experimental result for  $d_u\leq 1.25$. It is
observed that the scalar unparticle effect results in that the BR
becomes smaller than the SM result for the range $1.25\leq d_u\leq
1.70$. This is due to the mixing terms of the SM and the
unparticle contributions.

\begin{figure}[htb]
\vskip -3.0truein \centering \epsfxsize=6.8in
\leavevmode\epsffile{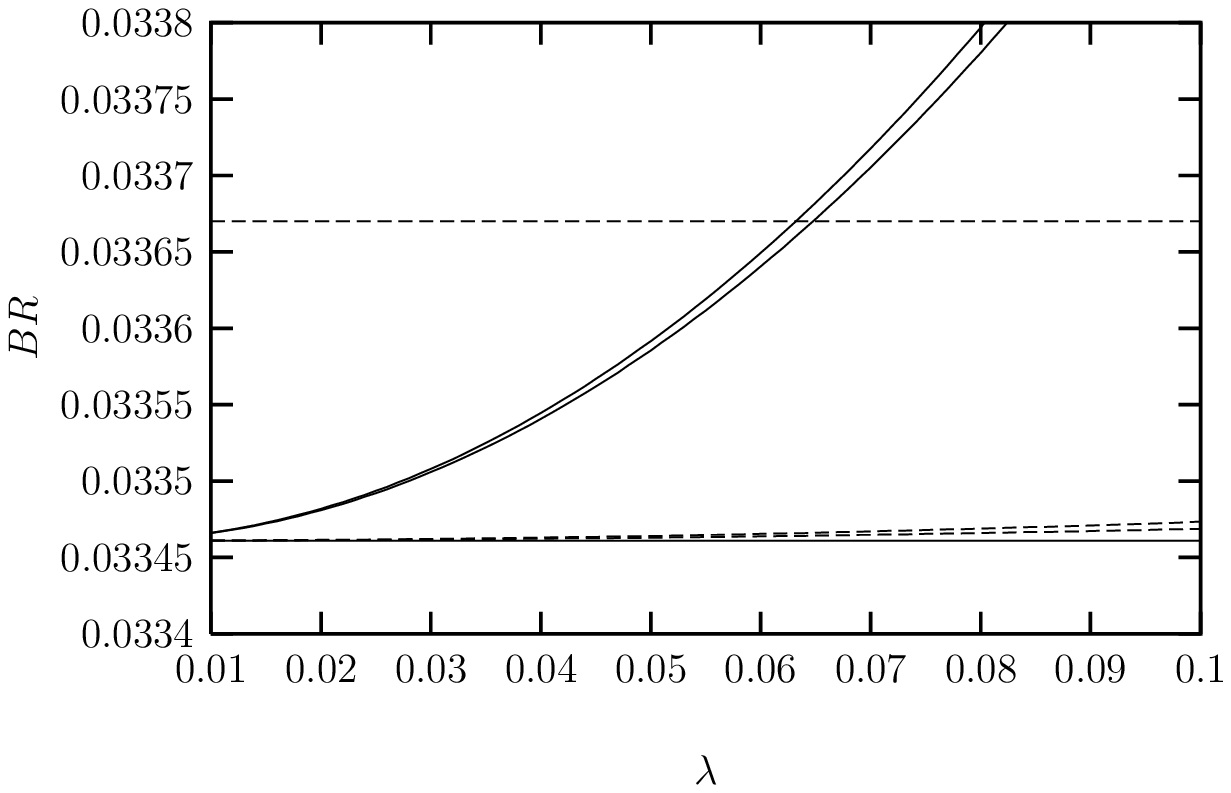} \vskip -3.0truein \caption[]{The
coupling $\lambda$ dependence of the BR $(Z\rightarrow e^+\,
e^-)$. The solid (dashed) straight line represents the SM
(experimental) BR and the lower-upper solid (dashed) curve
represents the BR with respect to $\lambda=\lambda_{ee}$ where
$\lambda_{\mu\mu}=10\,\lambda$, $\lambda_{\tau\tau}=100\,\lambda$,
$\lambda_{ij}=0.5\, \lambda$, $\lambda_0=\lambda_Z=10\,\lambda$,
for $\Lambda_u=10\, TeV$-$\Lambda_u=1.0\, TeV$ $d_u=1.01$
($d_u=1.1$) .} \label{Zeelam}
\end{figure}

\begin{figure}[htb]
\vskip -3.0truein \centering \epsfxsize=6.8in
\leavevmode\epsffile{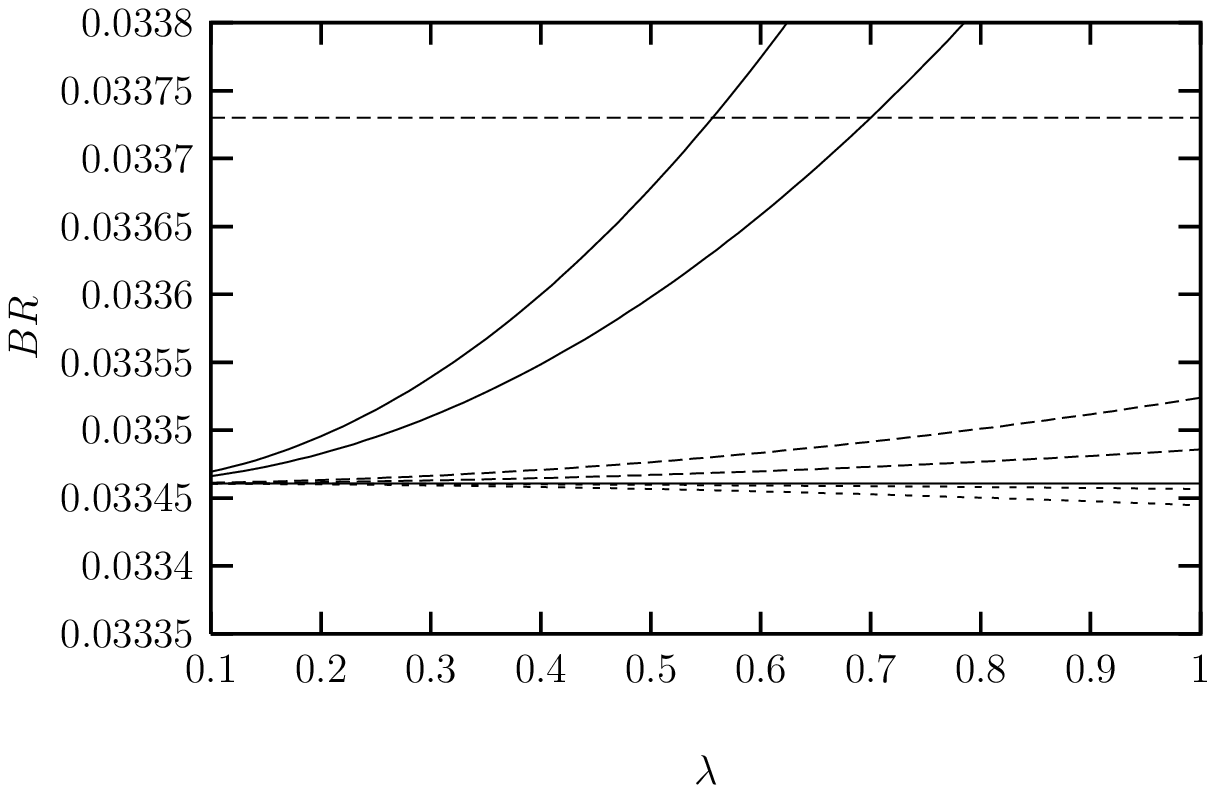} \vskip -3.0truein \caption[]{The
coupling $\lambda$ dependence of the BR $(Z\rightarrow \mu^+\,
\mu^-)$. The solid (dashed) straight line represents the SM
(experimental) BR and the lower-upper solid (the lower-upper
dashed, the lower-upper short dashed) curve represents the BR with
respect to $\lambda=\lambda_{\mu\mu}$ where $\lambda_{ee}=
0.1\,\lambda$, $\lambda_{\tau\tau}=10\,\lambda$,
$\lambda_{ij}=0.05\, \lambda$, $\lambda_0=\lambda_Z=\lambda$, for
$\Lambda_u=10\, TeV$-$\Lambda_u=1.0\ TeV$ $d_u=1.1$
($\Lambda_u=10\, TeV$-$\Lambda_u=1.0\, TeV$ $d_u=1.2$,
$\Lambda_u=1.0\, TeV$-$\Lambda_u=10\, TeV$ $d_u=1.3$) .}
\label{Zmumulam}
\end{figure}

\begin{figure}[htb]
\vskip -3.0truein \centering \epsfxsize=6.8in
\leavevmode\epsffile{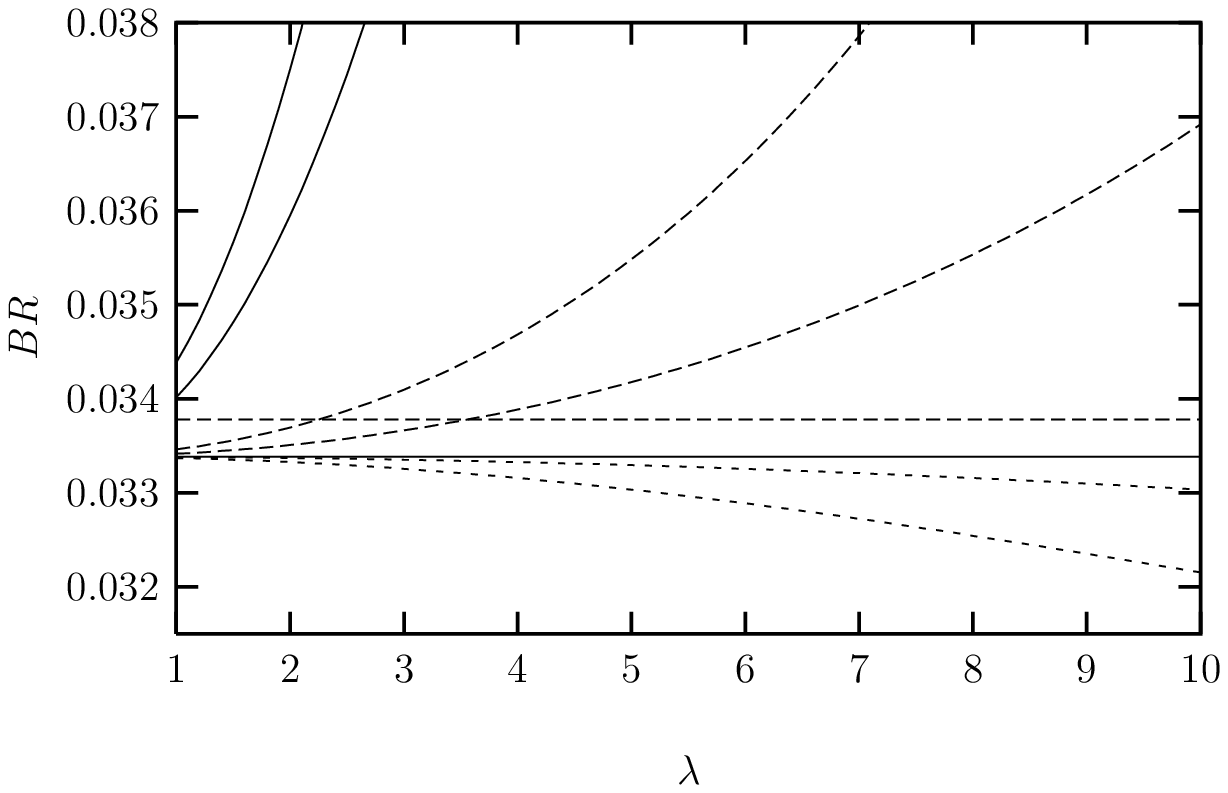} \vskip -3.0truein
\caption[]{The coupling $\lambda$ dependence of the BR
$(Z\rightarrow \tau^+\, \tau^-)$. The solid (dashed) straight line
represents the SM (experimental) BR and the lower-upper solid (the
lower-upper dashed, the lower-upper short dashed) curve represents
the BR with respect to $\lambda=\lambda_{\tau\tau}$ where
$\lambda_{ee}= 0.01\, \lambda$, $\lambda_{\mu\mu}=0.1\,\lambda$,
$\lambda_{ij}=0.005\, \lambda$, $\lambda_0=0.1\,\lambda$, for
$\Lambda_u=10\, TeV$-$\Lambda_u=1.0\, TeV$ $d_u=1.1$
($\Lambda_u=10\, TeV$-$\Lambda_u=1.0\, TeV$ $d_u=1.2$,
$\Lambda_u=1.0\, TeV$-$\Lambda_u=10\, TeV$ $d_u=1.3$) .}
\label{Ztautaulam}
\end{figure}

In Figs. \ref{Zeelam} (\ref{Zmumulam}, \ref{Ztautaulam}) we
present the BR $(Z\rightarrow e^+\, e^-)$ (BR $(Z\rightarrow
\mu^+\, \mu^-)$, BR $(Z\rightarrow \tau^+\, \tau^-)$) with respect
to the couplings $\lambda$, for different values of the scale
parameter $d_u$.  Here the solid (dashed) straight line represents
the QED corrected SM (the experimental) BR. In Fig.\ref{Zeelam}
the lower-upper solid (dashed) curves represent the BR with
respect to $\lambda=\lambda_{ee}$ where
$\lambda_{\mu\mu}=10\,\lambda$, $\lambda_{\tau\tau}=100\,\lambda$,
$\lambda_{ij}=0.5\, \lambda$, $\lambda_0=\lambda_Z=10\lambda$, for
$\Lambda_u=10\, TeV$-$\Lambda_u=1.0\, TeV$, $d_u=1.01$
($d_u=1.1$). It is observed that the experimental result is
reached for the numerical values of the scale parameter $d_u$ not
greater than $\sim 1.01$ for the coupling $\lambda > 0.065$. In
Fig.\ref{Zmumulam} the lower-upper solid (the lower-upper dashed,
the lower-upper short dashed) curves represents the BR with
respect to $\lambda=\lambda_{\mu\mu}$ where $\lambda_{ee}=
0.1\,\lambda$, $\lambda_{\tau\tau}=10\,\lambda$,
$\lambda_{ij}=0.05\, \lambda$, $\lambda_0=\lambda_Z=\lambda$, for
$\Lambda_u=10\, TeV$-$\Lambda_u=1.0\, TeV$ $d_u=1.1$
($\Lambda_u=10\, TeV$-$\Lambda_u=1.0\, TeV$ $d_u=1.2$,
$\Lambda_u=1.0\, TeV$-$\Lambda_u=10\, TeV$ $d_u=1.3$). The
experimental result is obtained for $d_u\sim 1.1$ and for the
coupling $\lambda > 0.5$ in the case that the energy scale is of
the order of $\Lambda_u=1.0\, TeV$. In Fig.\ref{Ztautaulam} the
lower-upper solid (the lower-upper dashed, the lower-upper short
dashed) curves represent the BR with respect to
$\lambda=\lambda_{\tau\tau}$ where $\lambda_{ee}= 0.01\, \lambda$,
$\lambda_{\mu\mu}=0.1\,\lambda$, $\lambda_{ij}=0.005\, \lambda$,
$\lambda_0=\lambda_Z=0.1\,\lambda$, for $\Lambda_u=10\,
TeV$-$\Lambda_u=1.0\, TeV$ $d_u=1.1$ ($\Lambda_u=10\,
TeV$-$\Lambda_u=1.0\, TeV$ $d_u=1.2$, $\Lambda_u=1.0\,
TeV$-$\Lambda_u=10\, TeV$ $d_u=1.3$). In this decay the
experimental result is obtained for $d_u\sim 1.2$ and for the
coupling $\lambda > 2.5$ in the case that the energy scale is of
the order of $\Lambda_u=1.0\, TeV$. For $d_u\sim 1.1$ the
experimental result is reached even for small couplings, $\lambda<
1.0$.

Now, for completeness, we would like to discuss the possibility of
mixing between unparticle and Higgs boson. The possible
interaction lagrangian which can induce such mixing \cite{Fox,
Feng} reads
\begin{eqnarray}
{\cal{L}}_{mix}= -\kappa_U\,H^\dagger\,H\, O_{U} \, ,
\label{lagrangianmix}
\end{eqnarray}
where $H$ is the Higgs field and $\kappa_U$ is the coupling with
mass dimension $2-d_U$. In the case that the Higgs field acquires
a non zero vacuum expectation value, the conformal symmetry of
unparticle sector is broken and the Higgs field mixes with the
unparticle operator $O_{U}$. Recently, the effect of the
considered mixing has been analyzed in detail \cite{Delgado1,
Delgado2}, based on the idea of deconstructed version of the
unparticle sector \cite{Stephanov}. The non zero vacuum
expectation value of the Higgs field drives the vacuum expectation
value for the infinite tower of scalars which construct the
unparticle operator and, therefore, the unparticle operator
$O_{U}$ develops non zero vacuum expectation value which results
in the conformal symmetry breaking. In these works, it has been
emphasized that, besides the conformal symmetry breaking in the
unparticle sector, the unparticle-Higgs mixing drives the possible
influence on the Higgs boson properties, like its mass and decay
width.

With the assumption that the conformal symmetry is broken at a
certain scale $\mu$, at least, the spectral density becomes
\begin{eqnarray}
|<0|O_U|P>|^2\,\,\rho(P^2)=A_{d_u}\,\theta(P^0)\,\theta(P^2-\mu^2)\,
(P^2-\mu^2)^{d_u-2} \, , \label{spectrdensty}
\end{eqnarray}
and this corresponds to remove modes with energy less than $\mu$.
We expect that the new form of the spectral density affects the
BRs of the Z boson decays under consideration since the unparticle
mediator which exists in the loops would be modified\footnote{This
modification needs more detailed calculation which we left for
future work.}.

As a summary, the LFC Z boson decays are sensitive to the
unparticle scaling dimension $d_u$ for its small values. The
experimental result of the BR is obtained for the parameter $d_u <
1.2$ for heavy lepton flavor output and the discrepancy between
QED corrected SM result and the experimental one can be explained
by the scalar unparicle effect. This may be a clue for the
existence of unparticles and informative in the determination of
the scaling parameter $d_u$. For light flavor output one needs to
choose the parameter $d_u$ near to one and, for the values of
$d_u$ which are slightly far from one, the discrepancy between QED
corrected SM result and the experimental can not be explained by
the unparticle contribution. Therefore, with the forthcoming more
accurate measurements of the decays under consideration,
especially the one with heavy lepton flavor output, it would be
possible to test the possible signals coming from the unparticle
physics
\newpage
\newpage
\newpage
\end{document}